# Temperature Secret in Bathtub: A Model of Temperature Distribution of Bathtub Based on Heat Conduction Equation

Yunfei Liu

## Summary

We utilize multidimensional heat conduction equation as well as heat transfer equation to establish a temperature distribution model of water in bathtub by solving partial differential equation. We solve the problems of the optimal strategy to add water and the optimal design of bathtub.

Firstly, we establish a cooling model of water surface based on Newton's law of cooling, simulating the heat exchanging process of two-dimension water surface between air and water surface. On this basis we consider the process of heat conduction, simulating the conduction process of the temperature in bathtub: If we do not add new heat source, the temperature of water body will reach the minimum point in 40 minutes.

Secondly, we discuss the situation of adding adequate hot water at once, establishing a one-dimensional heat conduction model, obtaining the stereogram and cross-section graph of temperature distribution. Then we add the influence of air cooling, perfecting our model. In the following step, we discuss the situation of steady heat source, establishing a one-dimension heat conducting model. By changing different values of the quantity of heat, we determine that the optimal heat input value is 80 Joules. In the meantime, we consider the situation of the heat dissipation of bathtub, obtaining that the optimal water velocity is $0.042 m/s$ by using heat equation. Utilizing these optimal parameters, we can maintain a certain temperature and save water.

Based on the final heat conduction model, we qualitatively analyze the factors such as human activities, gaining the optimal design plan of bathtub: The length is 1.5m, width is 0.6m, depth is 0.42m and the shape of bathtub is rounded angle rectangle. In addition, we also provide some suggestions about taking baths as well as adding water.

Eventually, we based on finite difference method and Pdetool in MATLAB, solving the heat conduction equation, verifying the stability of numerical solution. Furthermore, we discuss the advantages and disadvantages of the model, providing the improvement direction of the model.

# Temperature Secret in Bathtub

After a day's hard work, don't you want to enjoy a nice bath? However, normally, if you take a bath over a period of time, you will definitely feel cold. As we all know, it is troublesome to control the proper amount of added water and select the appropriate time when water should be added. Fortunately, we deliberately designed a kind of optimized bathtub, and hopefully, it will free you from the inconvenience of the normal bathtub mentioned above.

Based on the water temperature distribution model, we designed a special shaped bath model, making the water temperature well-proportioned in the bathtub, which can guarantee that you feel warm and comfortable all the time. However, due to the fact that the temperature will naturally decline, here we provide you with some suggestions on the affair of adding water that you should follow.

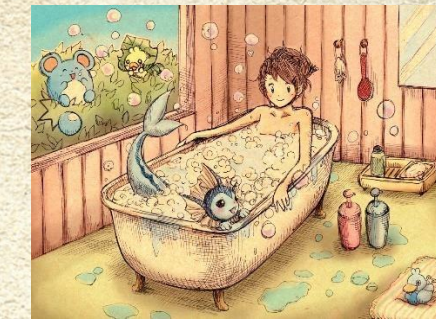

Firstly, you had better start adding water after 30 minutes' bath. The water faucet is specially designed. It is worth mentioning that hot water can be added only in a specific area, inevitably causing the temperature of the area near or under the water faucet being much higher. Therefore, we decrease the flux of the water and increase the temperature of the water, through which the area where temperature non-uniformly distributed can decline.

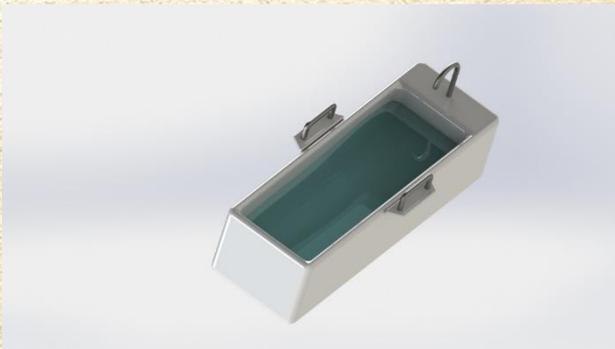

The water faucet is designed with constant flux as we stated. The only thing you need to do is to open the faucet. After around 20 minutes, the water temperature will be suitable steady. Furthermore, comparatively small amount of water flow rate can effectively help avoid the waste of water resources.

Besides, the height of the inner diameter of the bathtub fits like a glove, on the one hand, being capable of maintaining water temperature, and on the other hand, avoiding waste of water coming from people's entering into the bathtub.

At last, we sincerely hope you can fully enjoy a relaxing and comfortable bath.

# Contents

# 1 Introduction

## 1.1 Problem Review

It is a common thing in our lives to soak in a tub. Everyone wants to take it to reward himself after the whole day running around, but how to cool the bath water is always an uncomfortable problem. The topic, in the context of bubble bath, depicts a mathematical problem, namely using what kind of bathtub to have bubble bath. Specifically, it is closely related to the following two problems:

Firstly, determining the best strategy for bathing, that is when the hot water shall be added, how the hot water shall be added, and how much water shall be added, as well as the temperature of the water, which are the problems concerned. And these problems above are based on the two principles of maintaining the temperature constant and saving hot water to the greatest degree.

Secondly, determining the most appropriate bathtub shape. Just as mentioned in the topic, bathtub shape and human activities will also have an impact on the water temperature. Our purpose here is to gain the best bathtub model, to propose some suggestions to someone who baths in the meanwhile, and to fully enjoy the fun of bathing.

In order to solve the two problems, we must build a heat distribution model inside the tub, and the model is the key to solving problems correlated with bathing, and other problems are based on this. Therefore, we focus on the building of heat distribution model.

## 1.2 Problem Analysis

As stated in the previous section, we shall build a heat conduction model of bathtub. The water in the bathtub is fluid, so the model is not only related to the heat conduction under the three-dimensional circumstance, but also the heat convection of water. Besides, the heat exchange between water and the air of bathroom, between water and walls of bathtub, between human and water, which are all the factors shall be involved.

But the complex model not only brings a heavy burden to calculation, but also this nearly perfect model will mostly not be used in real bathing. Moreover, the essence of mathematical model lies in its abstract, namely grabbing the essential features of matters. Therefore, we have adopted the means of building models one by one and superposing abstractly to display the heat conduction model of bathtub based on different cases step by step. Hereof, we demonstrate the model in the graph below:

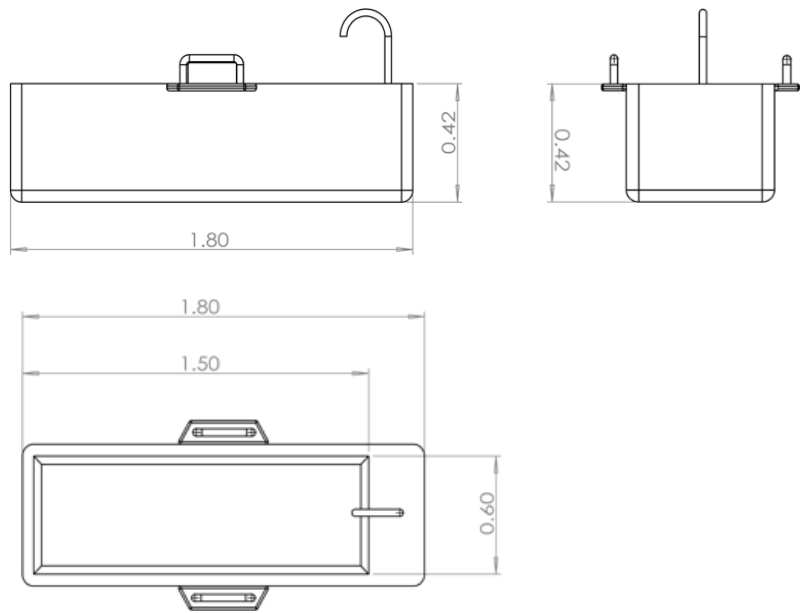

The parameters marked in the graph are the model parameter settings in our model.

As for thermodynamics problem, we mainly rely on the two basic equations, namely Newton's law of cooling and heat conduction equation to build the model. The process involves surface cooling, water heat conduction, air cooling, adding water locally, heat conduction of bathtubs, as well as adding water continuously, whose essence is the change of different forms and coefficient of heat conduction equations.

In the process of solving the model, we have tried different methods, eventually adopting the finite difference methods for partial differential equations to solve it and checking the stability and sensitivity of the model.

Finally, we put these factors together in an abstract way, and get the ultimate bathtub temperature distribution model, assessing the impact of various factors on the model, and putting forward some advice related to the strategy of adding bathing water and the design of bathtub based on it.

# 2 Assumptions

- The coefficient and specific heat of water, air, bathtub and human body are regarded as constants in a certain temperature range and the fixed pressure conditions.

- The effect of heat conduction of the water itself shall be considered, and the flow heat convection effect will not be taken into account. Because of the low water temperature, less amount of water, the heat convection effect is not so obvious, thus the water can be treated as rigid one, which only involves heat convection.

- The impact of time on the adding water locally shall not be considered. We can assume like this, for adding water locally is actually to give partial differential equations initial value, in real cases, which equals adding hot water into the bathtub instantly.

- The air temperature of bathroom only relates to the heat exchange between water surface and air, which will not change with other variables. For the process of bathing, the main heat exchange object of air is water surface not others like ceilings.

- The data we have collected is reliable.

- Assuming the person who baths will not lower the water temperature deliberately, that is he hopes to maintain the water temperature, which is in line with our common sense.

- Bubble baths will only affect contact area between water surface and air. In fact, bubble bath plays a certain role of maintaining temperature and thermal insulation, equivalent to the reduction of contact area.

- The comfort of bathing is only concerned with water temperature, having nothing to do with other factors.

# 3 Symbol Description

| Symbol | Explanation |
| --- | --- |
| $T$ | The temperature of every point in the bathtub |
| $T_c$ | The outside temperature of the bathtub is namely the air temperature transferring heat with the water body. |
| $T_0$ | The initial temperature of the water |
| $\rho$ | The density of every kind of material (water and air). |
| $C$ | The specific heat capacity of every kind of material (water and air). |
| $k$ | The coefficient of heat conduction of every kind of material |
| $h$ | The height of the water body in the bathtub. |
| $A$ | The surface area of the water body. |
| $V$ | The volume of the water in the bathtub. |
| $h_{air}$ | The coefficient of heat convection of the influence between air and the water surface. |
| $f$ | The heat source is added into the water in the bathtub. |
| $Q$ | The quantity of heat. The unit is Joule. |

# 4 Models

As the former description, the best model is easy and clear for readers to understand. Therefore, we establish the model considering the factors such as the cooling of water surface, the heat conduction of water body, the heat dissipation of air, locally adding water, the heat conduction of bathtub and continuously adding water. Eventually, we integrate these models then obtain the final model and also optimize the final model in other aspects.

Moreover, we reduce the dimensions for all of our models, thus our models are established on the basis of one-dimension or two-dimension. Because low dimensional models are actually the slices of high dimensional models, low dimensional models further possess the general representation, also being more convenient for persons to check and understand.

## 4.1 Cooling Model of the Water Surface

Here, we first consider the cooling model of the water surface. Obviously, due to the water surface contacts with air, the water surface temperature will decline. Hence we consider the cooling situation of the two-dimensional surface based on Newton's law of cooling. The expression of Newton's law of cooling is

$$\Delta t = t - t_0 = T \ln \frac{(T_o - T_c)}{(T - T_c)}$$

Where $\Delta t$ is the temperature variation of the entirety at one point. There out, we can deduce the convective heat transfer equation as follows:

$$Q = h_{air} A (T - T_c)$$

Where Q is the exchange quantity of heat, $h_{air}$ is the coefficient of heat convection, A is the unit area of contact.

This model can assist us to intuitively observe the temperature variation. However, we cannot obtain the temperature variation of the water body in bathtub. As a result, we based on the heat conduction equation establish the following model of heat conduction together with heat dissipation of the water surface.

## 4.2 Heat Conduction Model of Water Body and Heat Dissipation Model of Air

On the basis of the heat dissipation model, we now simultaneously consider the heat conduction coupled with the heat dissipation of water surface, gaining a comprehensive model. Here, we get the inspiration from the optimal chocolate cake bakeware design (MCM problem A, 2013), quoting the two-dimension expression of heat conduction equation as follows:

$$\rho c \frac{\partial u}{\partial t} = h_{air} \frac{\Delta A}{\Delta V} (T - T_c)$$

Where $\Delta V$ is the unit volume of water body, k is the coefficient of heat conduction of water.

Three-dimension graphs are difficult to understand and not intuitive. The three-dimension heat distribution graph can derive from the two-dimension graph. Thereupon we set different time, obtain a set of two-dimension heat distribution graphs of water surface as follows:

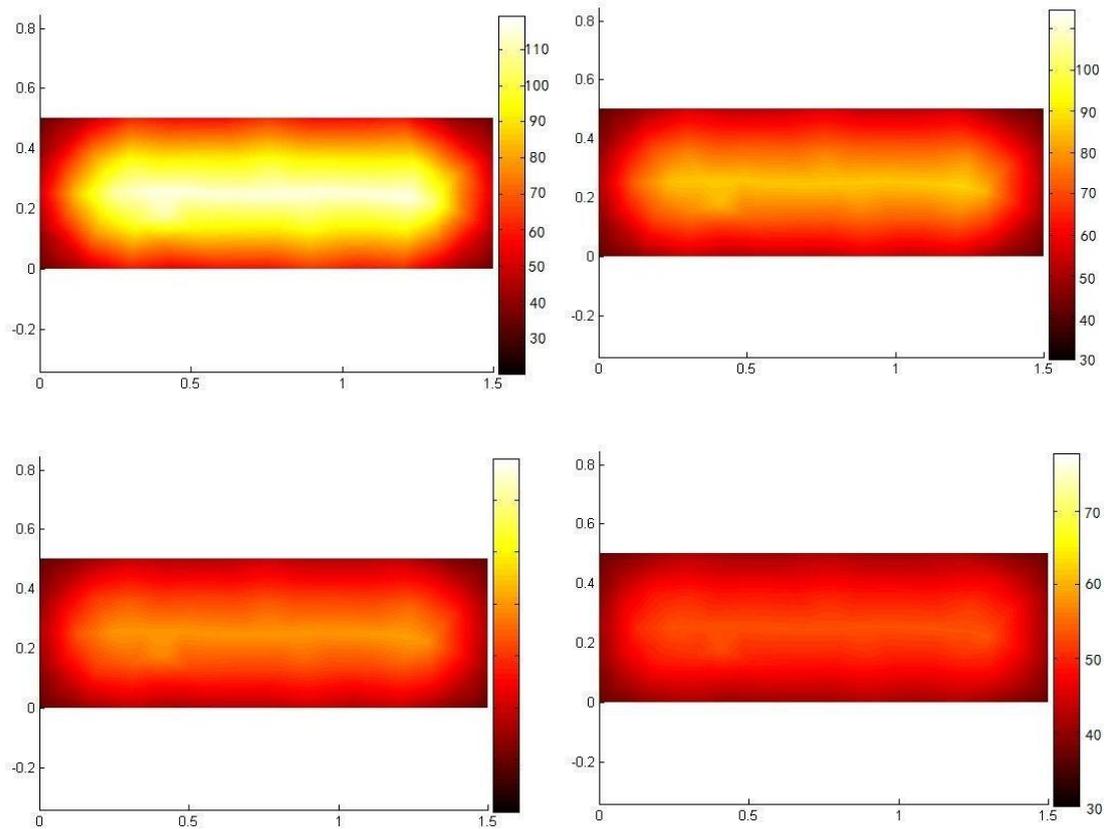

Figure 1. The temperature distribution graph of two-dimension heat conduction of water body at different time

Here the horizontal axis shows the length of bathtub, vertical axis shows the width of bathtub, color shows the temperature of bathtub. We set ten minutes as unit time, $88.4F°$ as initial temperature.

From the four graphs, we can see that with time the temperature of water surface gradually declines. In the end, the temperature reaches the minimum (about $47.6F°$). This phenomenon conforms to common sense.

Please pay attention: In this model, we do not add hot water to bathtub. That is to say, there is no influence of the external stable heat source Q. The factor considered is the heat conduction of water body as well as the heat convection between the water surface and air. Obviously, if we do not add hot water, the person who has a bath will feel cold in thirty to forty minutes, then the person himself must add hot water to keep the constant temperature of bathtub.

Therefore, in the following step we will start to discuss the adding water models.

## 4.3 Heat Conduction Model of Locally Adding Water

As mentioned earlier, we cannot gain the temperature distribution of all parts of bathtub from the above models. Thus, we consider the heat conduction model of bathtub based on heat conduction.

The heat conduction is the heat transfer phenomenon when there is no macroscopic movement in the medium. This phenomenon can occur in solid, liquid and gas. But strictly speaking, the pure heat conduction only exists in solid. Because in fluid, the liquid itself will generate natural convection due to the density difference caused by the temperature gradient, namely the heat convection. So, the heat conduction and the heat convection in liquid coexist. (Ozisik, M. Necati, 1993)

However, in order to simplify the model, we just regard the water in the bathtub as solid rather than liquid. Meanwhile we merely consider the one-dimension heat conduction. In this way, we can build a one-dimensional model of the heat conduction bathtub based on the heat conduction equation. The expression is

$$\frac{\partial T}{\partial t} - \frac{k}{c\rho}\frac{\partial^2 T}{\partial x^2} = 0$$

In this model, we assume that the initial heat (hot water) is suddenly added, and the excess water is also suddenly drained, namely we do not consider the time influence of the adding and draining process. The water in bathtub possesses a certain initial temperature. As a result, we obtain the graph as follows:

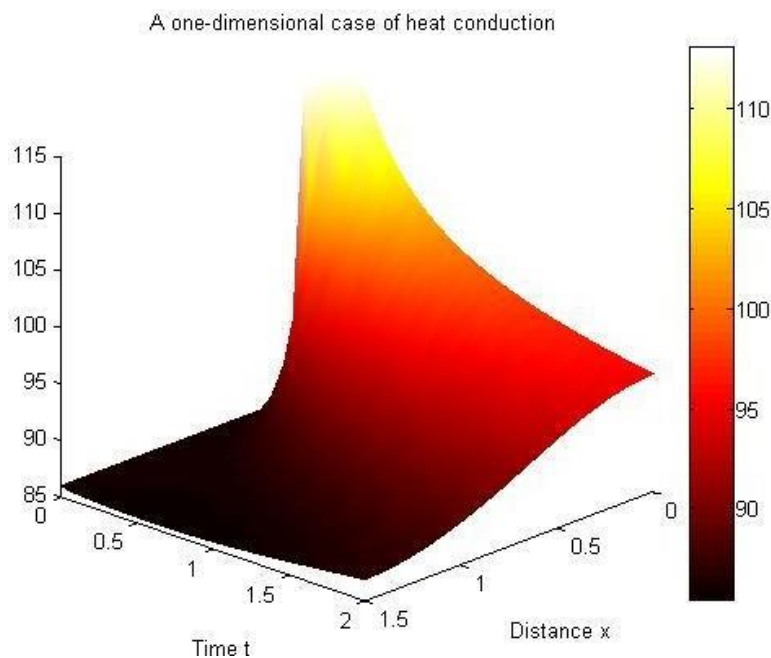

Figure 2. The sketch graph of heat conduction model by locally adding water

Here we locally add hot water, that is adding $113F°$ hot water in $0<x<0.3$ section. The rest water temperature maintains $86F°$. Visibly, the initial heat (hot water) will quickly decline, and the bulk temperature will escalate, eventually reaching a steady state, namely the water temperature maintains constant for the reason that we ignore the heat dissipation of air.

Meantime in order to be convenient to understand, we also draw the cross-section graph of the temperature distribution under different time nodes. We properly adjust partial parameters, gaining the following graph:

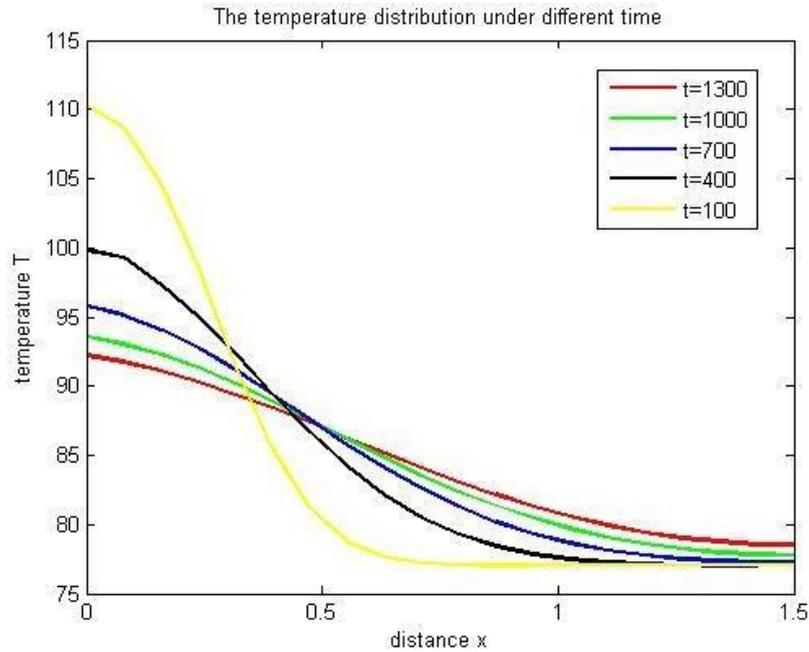

Figure 3. The slice graph of heat conduction model by locally adding water at different time

Obviously, when t is 100, the temperature decreases most quickly in one-dimension space. When t is 1300, the temperature decreases and finally tends to be flat. The final temperature is slightly higher than $77F°$, which is caused by adding a certain amount of hot water. This phenomenon verifies energy conservation law in another aspect as well.

Nonetheless, this model still cannot perfectly simulate the actual temperature distribution model, therefore we based on this model consider the heat dissipation model of air to the water surface.

## 4.4 Heat Conduction and Dissipation Model of Locally Adding Water

This model is also established in one-dimension space, and also can be extended into three-dimension space with the same principle. Based on the former basis we add the influence of the temperature distribution caused by heat convection between the water surface and air. We improve the equation mentioned in 4.1.3, obtaining the following equation:

$$c\rho \frac{\partial u}{\partial t} = k \times \nabla^2 T + h_{air} \frac{\Delta A}{\Delta V}(T - T_c)$$

This equation can be seen as the integration of two former models. As a result, we gain the following heat distribution graph. The parameter selection as well as adding water method is the same as 4.1.3.

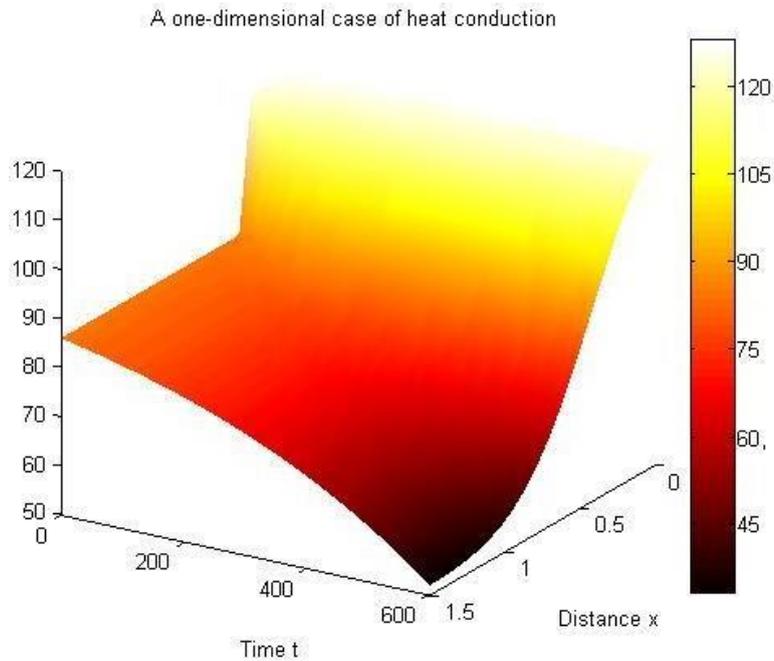

Figure 4. The sketch graph of heat conduction and heat dissipation model by locally adding water

We can observe that the final temperature of this model does not approach a certain value, but continuously decreases. It is easy to understand: By reason of merely adding the initial source of heat, the water body continuously dissipates heat to air. Whereupon the final temperature continuously decreases. However, temperature declining under 32 $F^\circ$ makes no sense. Therefore, this stereogram can just be seen as a sketch graph.

In the same way, we also intercept the heat distribution graph of different temperature as follows:

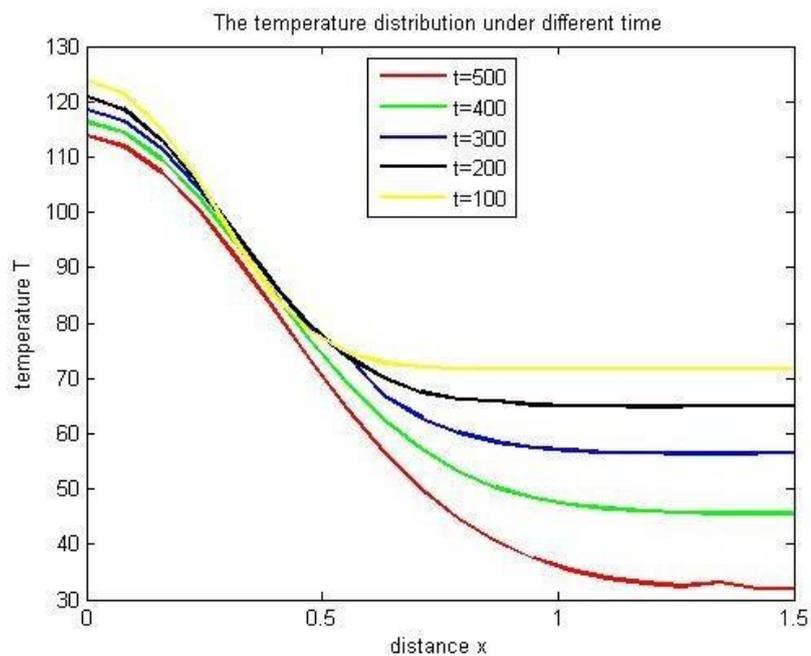

Figure 5. The slice graph of heat conduction and heat dissipation model by locally adding water at different time

Different from the segmented heat distribution graph in 4.1.3, we find the temperature of the heat distribution graph tends to different values at different times because in this model we add the heat dissipation factors. In the meantime, we can observe that with time going, the temperature is lower. This phenomenon confirms common sense.

To summarize, we can get a more practical model through comparing the two former models. Here we redraw the temperature distribution graphs of two former models when t is 400 as follows:

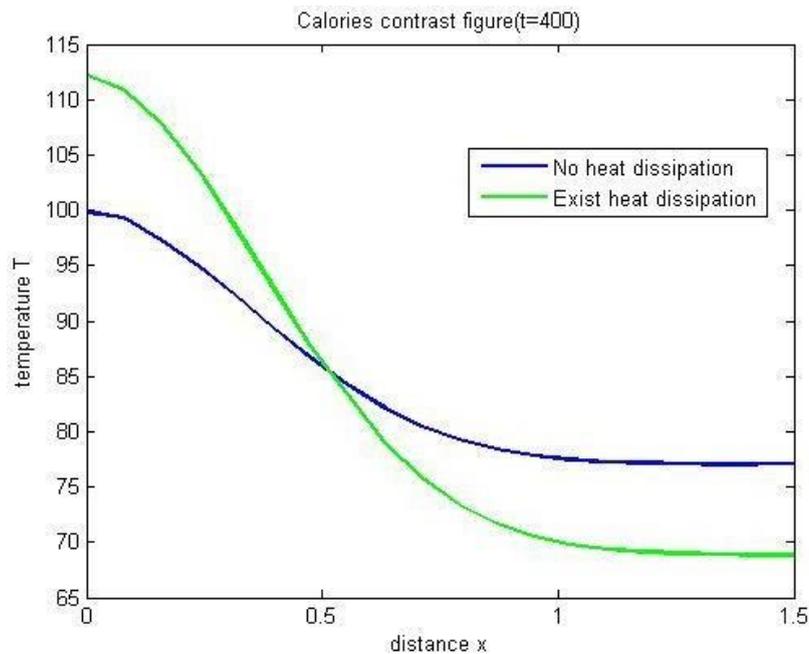

Figure 6. The comparison graph of temperature distribution of two models

We can draw the conclusion: At the same time, air cooling will result in the temperature of water body declining more rapidly and reaching a lower value.

The problem mentioned that the person would open the hot water faucet to maintain a constant quantity of heat to enjoy a more comfortable bath. Therefore, we will consider adding a constant source of heat to the model next.

## 4.5 Continuous Water Model

Due to the request of the problem, we add a constant source of heat and stipulate that this heat source is added on one side of the one-dimension model. Besides, the initial conditions are different from the former models. This model assumes that the whole temperature of the water body is under $100F°$, instead of locally adding water. In this way, we utilize the equation:

$$c\rho \frac{\partial T}{\partial t} = k \times \nabla^2 T + h_{air} \frac{\Delta A}{\Delta V}(T - T_c) + Q$$

to gain a continuous water model based on the heat conduction equation. We continuously change the value of Q to find the suitable temperature condition. We assume that the optimum temperature is about 100 to102F°, and the best value of Q is about 80J. The temperature distribution graph is

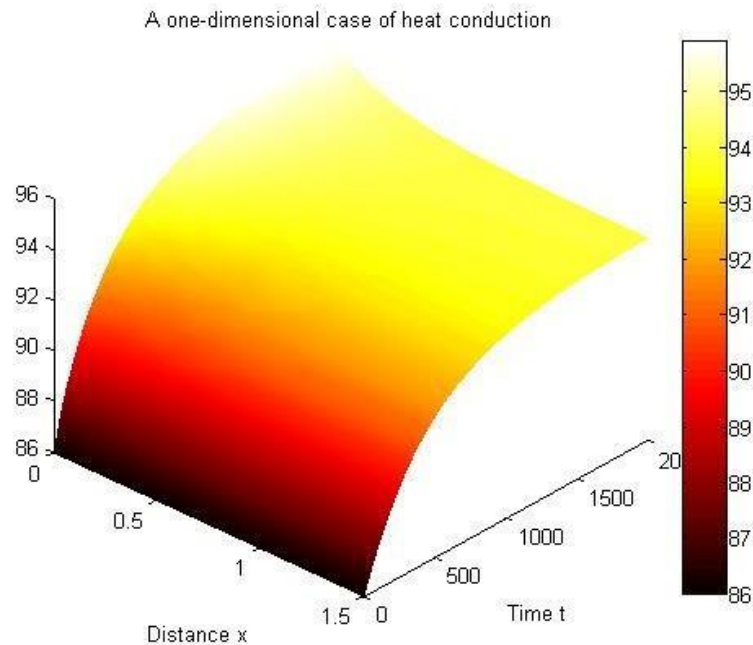

Figure 7. The sketch graph of continuously adding water model

We can observe that under the premise of ignoring water saving, the final temperature of bath water maintains in a constant state and rapidly rises, maintaining constant in 1500 seconds. Hence, continuously adding water is necessary and effective. At the same time, here we also give the final temperature under different conditions of heat form as follows:

| Q(J): | 70 | 75 | 80 | 85 | 90 |
|---|---|---|---|---|---|
| Temperature($F°$): | 99.89 | 100.78 | 101.71 | 102.60 | 103.51 |

Table 1. The steady-state temperature of different values of quantity of heat

However we cannot obtain some specific data including water-carrying capacity from this model. We also ignore the heat dissipating capacity of the wall of bathtub. We construct the following model of the heat conduction bathtub in order to shorten the differences of actual situations.

## 4.6 Heat Conduction Model of the Bathtub

Figure 8. The water column

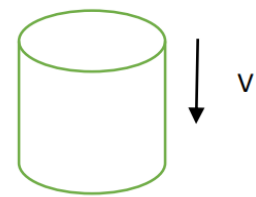

Due to the continuous cooling of bathtub, heat dissipates in unit time. From the definition of the heat conductivity:

$$k_x = -\frac{q''_x}{\frac{\partial T}{\partial x}}$$

Where $k_x$ is the heat conductivity of medium, q is the density of heat flow in X direction, $\frac{\partial T}{\partial x}$ is the temperature gradient in X direction, in terms of the isotropic homogeneous medium (such as the wall of bathtub) we can obtain

$$q = -\frac{k_x \Delta T S}{d}$$

Where q is the heating unit time, $\Delta T$ is the temperature difference between the inside and outside wall, S is the contact area, d is the width of the wall. We take a rectangular bathtub that is 1.5 meters long, 0.6 meter wide, 0.42 meter deep as an example, plugging in the relative data:

$$S = 1.908 m^2 \qquad q_1 = -18.126 J$$

From the model 4.1.5, in order to maintain the constant water temperature in 77.6°F or so, the system needs the heat (q = 80J) in unit time to counteract the heat loss in air. The heat $q_2$ brought by the water faucet confirms to

$$q = q_1 + q_2$$

Then

$$q_2 = q - q_1 = 98.126 J$$

The heat brought by hot water in unit time is

$$q_2 = \rho c v S \Delta T = 98.126 J$$

Where $\rho$ is the density of water, $c$ is the specific heat capacity of water, $S$ is the cross-sectional area of the water column, $\Delta T$ is the reduced temperature of hot water. We plug in the data and gain

$$v = \frac{q_2}{\rho c \Delta T S} = 0.042 m/s$$

As a result, we obtained the heat conduction model of the wall of bathtub, and figured out the water velocity of the constant source of heat is about $0.042 m/s$.

Compared to common sense, this water velocity is relatively credible. We will describe the specific model test in the latter parts.

## 4.7 Model Implement

This model is actually a complicated four-dimension heat conduction equation. To solve this equation, we utilize four different methods:

- The toolbox Pdetool to solve the partial differential equation in MATLAB.
- The specific function Pdepe to solve the partial differential equation in MATLAB.
- Transform the partial differential equation into finite element and finite difference equation.
- Utilize Fourier change to transform partial differential equation to solve linear partial differential equation, then utilize inverse Fourier change to get the result.

Pdepe function possesses relatively great universality, but is complex for us to program; PDE toolbox supplies GUI interface, can assist us to solve specific PDE problems, but possesses relatively great limitation. These methods can both assist us to solve the two-dimension heat conduction equation but possess no ability to solve three-dimension equations.

In terms of Fourier change, it is tremendously complex and insignificant to gain the symbolic solution. Therefore, we do not utilize this method.

As a result, we mainly utilize finite elements as well as finite difference equations to solve. The main idea of finite elements is to incise and divide its solution space, simplifying the solving process. The general three-dimension heat conduction equation is

$$c\rho\frac{\partial T}{\partial t} - k \times \nabla^2 T + h_{air}\frac{\Delta A}{\Delta V}(T - T_c) = f$$

In terms of the approximate calculation of finite difference method, we use $T(t^+)$ to denote T (x, y, z, t+$\Delta t$) while use $T(t^-)$ to denote T (x, y, z, t-$\Delta t$). Then we can gain the approximate results as follows:

$$\frac{\partial T}{\partial t} = \frac{T(t^+) - T}{\Delta t}$$

$$\frac{\partial^2 T}{\partial x^2} = \frac{T(x^+) - 2T + T(x^-)}{\Delta x^2}$$

Similarly, we can deduce the other approximate results, plug these approximate expressions into partial differential equation and get the equation.

$$c\rho\frac{T(t^+) - T}{\Delta t} - k\left(\frac{T(x^+) - 2T + T(x^-)}{\Delta x^2} + \frac{T(y^+) - 2T + T(y^-)}{\Delta y^2}\right.$$

$$\left. + \frac{T(z^+) - 2T + T(z^-)}{\Delta z^2}\right) + h_{air}\frac{\Delta A}{\Delta V}(T - T_c) - f = 0$$

Then we can obtain the following recurrence equation:

$$T(t^+) = T\left(1 - \frac{\Delta t}{c\rho}\left(2k\left(\frac{1}{\Delta x^2} + \frac{1}{\Delta y^2} + \frac{1}{\Delta z^2}\right) + h_{air}\frac{\Delta A}{\Delta V}\right)\right)$$

$$+ \frac{\Delta t}{c\rho}k\left(\frac{T(x^+) + T(x^-)}{\Delta x^2} + \frac{T(y^+) + T(y^-)}{\Delta y^2} + \frac{T(z^+) + T(z^-)}{\Delta z^2}\right)$$

$$+ \frac{\Delta t}{c\rho}\left(h_{air}\frac{\Delta A}{\Delta V}T_c + f\right)$$

We can transform this recurrence equation into a large sparse matrix and solve the equation by programming in MATLAB. The block graph to solve the general partial differential equation is

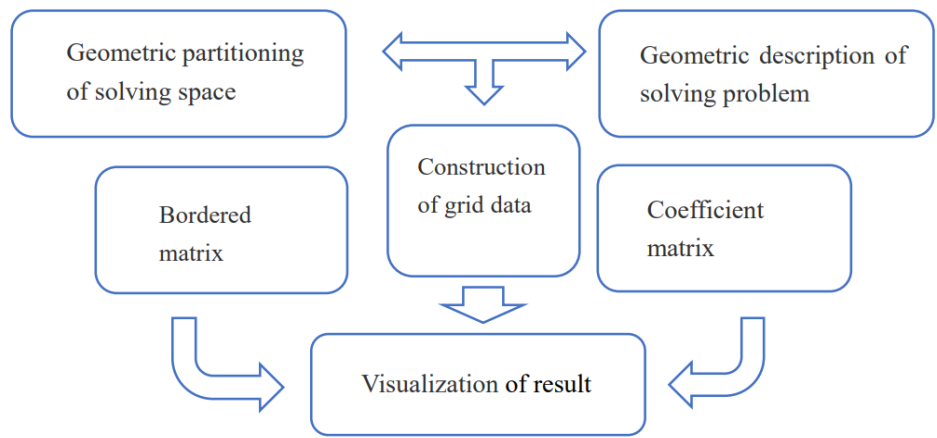

Thus, we can obtain the numerical solution of this model as well as the expression form of the result of visualization.

# 5 Examination

## 5.1 Influence of Other Factors on the Model

### 5.1.1 Influence of Human on the Model

- The influence of man's entering into the bathtub on the temperature distribution model: We suppose man will choose to enter into the bathtub from the place where the temperature is closest to his body surface temperature. At the same time, people's body surface temperature, being constant temperature source, will not change the temperature distribution of water surface at the isothermal place, but will partially drain away water, making the gage height rise. Therefore, in the temperature distribution model, we will take into consideration that gage height caused by people's entering into the bathtub.
- The influence of man's moving in the bathtub on the temperature distribution model: People's moving under the water will increase the convection of water, which means increasing water's thermal diffusivity k and thus making the water temperature much faster reach steady state.
- The influence of man's moving at the water surface on the temperature distribution model: people's moving at the water surface will increase the contact area between water and air, namely, increasing water surface area $\Delta A$ and making heat dissipation at a higher speed.

### 5.1.2 Influence of Bubble Bath on the Model

Bubble bath, compared to traditional water bath, is a bath which can make you feel more relaxed and comfortable. In our model, we only consider that the bubble bath affects the surface area in contact with air.

The effects the bubble bath makes are opposite to those of people in the water. The bubble will prevent water in contact with air, which reduces the surface area $\Delta A$ of water and slows the water cooling so as to achieve the effect of heat insulation.

## 5.2 Stability Test of the Model

We apply difference form to solve partial differential equation, while. Therefore, here we can apply Von Neumann stability analysis. To ensure the stability of numerical solution,

$\Delta x, \Delta y, \Delta z, \Delta t$ must match the following equation:

$$k\Delta t \left(\frac{1}{\Delta x^2} + \frac{1}{\Delta y^2} + \frac{1}{\Delta z^2}\right) \leq \frac{1}{2}$$

Therefore, for a given time step $\Delta t$, the selection of the spatial mesh size is within certain limits. Having passed the Von Neumann stability analysis, the selection of the spatial mesh size in the model is considerably suitable, occupying a better stability.

## 5.4 Sensitivity Analysis

The sensitivity analysis of the model is divided into the following two aspects. One is the effect of bathtub height h on models and the other is the influence of the heat transfer coefficient of water on the model. We have changed different parameters, getting the following two tables.

| Depth h(m) | 0.80 | 0.67 | 0.57 | 0.51 | 0.42 | 0.36 |
|---|---|---|---|---|---|---|
| Temperature($F°$) | 106.542 | 104.698 | 102.996 | 101.746 | 101.212 | 99.24 |

Table 2. The steady-state temperature of different depth

From the two tables we can clearly see that with the depth of bathtub deepened, the final temperature is getting higher and higher. We've stipulated body's best water temperature range, and we used the tub model with only 0.42m deep in order to facilitate the calculation.

| k(W/(m·K)) | 0.2 | 0.4 | 0.6 | 0.8 | 1 | 1.2 |
|---|---|---|---|---|---|---|
| Temperature($F°$) | 102.758 | 102.02 | 101.696 | 101.5 | 101.372 | 101.264 |

Table 3. The steady-state temperature of different coefficient of heat conduction

We can see from this table that temperature will be slightly increased with the reduction of the heat transfer coefficient of water. This is because the heat exchange between water and air becomes weak. In addition, we also see the temperature change is not large. It verifies our basic assumptions. At the same time, it indicates that the perturbation of individual parameters does not affect the model.

To sum up, our model is robust. It can be used to simulate a real situation of heat conduction bathtub.

# 6 Conclusion

## 6.1 Optimal Strategy for Adding Water

Due to the extremely unbalanced distribution of the temperature of inner space with instantly part water adding, we gave up this kind of water-adding strategy. Instead, we selected the continuous water-adding strategy, using the faucet whose diameters are 10MM. It makes the flow rate of $113F°$ hot water reach about 0.042m/s. In this way, not only the water won't be wasted too much but also the temperature of most areas in the space (except the nearby tap) will keep in an appropriate temperature $101F°$ of human body.

And the time of water adding is best when you add water into bathtub from the outset so that the temperature in bathtub will be kept in its best.

## 6.2 Optimal Design of Bathtub

From the two-dimensional graph we can observe that the temperature in four corners of the bathtub rapidly declines. Therefore, the optimal bathtub should possess rounded angle to enable the temperature in bathtub to be more balanced.

In addition, the area of bathtub is larger, the quantity of heat evaporating in unit time is larger. Therefore, the volume of bathtub should be slightly larger than the volume of human. That is to say under the premise of guaranteeing the comfort degree of bathing, the bathtub is smaller the effect is better. For the reason that a relatively small but effective bathtub can avoid heat loss.

By investigating the price of bathtubs on the market, we obtain the minimum length of bathtub and width b are respectively 1.5 meters and 0.6 meter.

In the heat conduction model of water body and heat dissipation model of air, due to the parameter:

$$\frac{1}{h_w} = \frac{\Delta A}{\Delta V}$$

excises in the equation:

$$\rho c \frac{\partial u}{\partial t} = h_{air} \frac{\Delta A}{\Delta V}(T - T_c)$$

We continuously change the h value of bathtub, maintaining the final steady temperature near the preference temperature of human body. Eventually we gain a relatively appropriate $h_w$ value: 0.34 meter.

Considering the water loss caused by a specific person entering the bathtub, in order to save water, we add the height of bathtub. The volume of human is generally about

$$V = 70 dm^3$$

Thus, a human entering the bathtub will mostly make the water level rise

$$\Delta h = \frac{V}{a} = 0.078m$$

Therefore, we set the depth of bathtub: b

$$h = h_w + \Delta h \approx 0.42m$$

To sum up, the optimal bathtub we design is 1.5 meters long, 0.6 meter wide, 0.42 meter deep. The shape is rounded angle rectangle.

Meanwhile, we also search the Kohler bathtub whose type is K-1150-LA-0 Bancroft 5-Foot Bath on Amazon's website. Kohler bathtub is 1.52 meters long, 0.5 meter wide, 0.51 meter deep, which sufficiently illustrates that our model confirms reality and is reasonable for bathtub design.

## 6.3 Other Suggestions

- In order to maintain the water temperature and get more balanced distribution of temperature in the bathtub, we should try our best to make our movements under water when we are in the bath instead of staying near or above the surface of the water.

- Meanwhile, using bubble bath can also achieve the objective of insulation.

- You can add a heater in the bathroom so that the temperature in the bathroom remains unchanged. And it effectively slows the water cooling.

# 7 Extension

## 7.1 Advantages of the Model

- We've used multiple models to describe the distribution of water temperature. At last, we synthetically optimized every model and got the final result.

- We've presented two different strategies of water adding. We compared and interpreted the results of the two models.

- We've analyzed the human activities and the effects of bubble bath on temperature distribution model and finally after the changes of the parameters the results of models are given.

- We've qualitatively analyzed the shape of the bathtub and quantitatively got the specifications as well as the continuous water-adding rate of faucet.

## 7.2 Disadvantages of the Model

- We replaced the temperature distribution model in three-dimensional space with the model in just one and two-dimension space.

- We didn't quantize human movements and the effects of bubble baths.

- We just studied the bathtub model which is similar to rectangle, but we didn't analyze the temperature distribution in other shapes.

## 7.3 Applicable Conditions of the Model

As mentioned above, we only made a detailed discussion of cases in two-dimension or even one-dimension space. The three-dimensional space does not apply in it. But we can get the case of three-dimensional space by expanding and extending reasonably. As you can imagine, by superposing heat distribution graphs of one-dimensional space together, you can get graph of three-dimensional space.

This model has a strong scalability because we made it on the basis of the heat conduction equation. It can be applied to lots of heat conduction fields such as heat conduction of steel, fire spreading and so on.

## 7.3 Improvement Directions of the Model

The main drawback of the model is the absence of a heat distribution graph of three-dimensional space. We can use the finite difference method mentioned above to program and solve, and we can also try using Krylov subspace methods of three-dimensional heat conduction equation to solve in the reference. (Dandan Li, 2010)

Meanwhile, we should also consider the heat exchange in flows of bathtub, which can be macroscopically given a little by computer simulation of cellular automata for three-dimensional space. In addition, it is a bit single of our evaluation criterions towards people's comfort levels of bath. Some factors affecting baths can be considered such as the degree of turbulence effects.